\documentclass[12pt]{iopart}

\usepackage{amssymb}
\usepackage{graphicx}
\usepackage{dcolumn}
\usepackage{bm}
\usepackage{enumitem}
\graphicspath{{figures/}} 

\begin{document}

\title[An analysis of SPR via Monte Carlo simulations]{Ligand-receptor binding kinetics 
       in surface plasmon resonance cells: A Monte Carlo analysis}

\author{Jacob Carroll$^{1}$, Matthew Raum$^{2}$, Kimberly Forsten-Williams$^{3}$ and  
        Uwe C. T{\"a}uber$^{1}$}

\address{$^{1}$ Department of Physics \& Center for Soft Matter and Biological Physics 
        (MC 0435), Robeson Hall, 850 West Campus Drive, Virginia Tech, Blacksburg, 
		VA 24061, USA}

\address{$^{2}$ Baker Hughes, 2851 Commerce Street, Blacksburg, VA 24060, USA}

\address{$^{3}$ Biomedical Engineering, Duquesne University, Pittsburgh, PA, USA}

\ead{jac21934@vt.edu, tauber@vt.edu}


\begin{abstract}
Surface plasmon resonance (SPR) chips are widely used to measure association and 
dissociation rates for the binding kinetics between two species of chemicals, e.g., cell 
receptors and ligands. It is commonly assumed that ligands are spatially well mixed in 
the SPR region, and hence a mean-field rate equation description is appropriate. This 
approximation however ignores the spatial fluctuations as well as temporal correlations 
induced by multiple local rebinding events, which become prominent for slow diffusion 
rates and high binding affinities. We report detailed Monte Carlo simulations of ligand 
binding kinetics in an SPR cell subject to laminar flow. We extract the binding and 
dissociation rates by means of the techniques frequently employed in experimental analysis 
that are motivated by the mean-field approximation. We find major discrepancies in a wide 
parameter regime between the thus extracted rates and the known input simulation values. 
These results underscore the crucial quantitative importance of spatio-temporal 
correlations in binary reaction kinetics in SPR cell geometries, and demonstrate the 
failure of a mean-field analysis of SPR cells in the regime of high Damk{\"o}hler number 
$Da>0.1$, where the spatio-temporal correlations due to diffusive transport and 
ligand-receptor rebinding events dominate the dynamics of SPR systems.
\end{abstract}


\vspace{2pc}

\submitto{\PB -- \today}

\maketitle

\section{Introduction}
The accurate measurement of the reaction rates between different species of 
chemicals is a crucial component in the process of understanding and 
manipulating the biochemical processes which perpetuate or extinguish life 
\cite{Biochem-Book, Voet}. 

A common method of measuring these rates is via surface 
plasmon resonance (SPR) \cite{SPR-Book, Phizicky}. 
SPR allows the binding dynamics between two species of 
chemicals to be measured in real time, and is performed 
by binding one of the two chemical species to a substrate 
(the receptor species), and then measuring the change in 
index of refraction as the other chemical species (the 
ligand species) flows over the substrate and the two chemicals 
interact \cite{Rich-1, Rich-2, Rich-3}. See Fig. \ref{fig:spr_cell}
for a schematic of the experimental setup. 

Ideally, the data from this experiment allows for the easy 
extraction of the binding and unbinding rates. However, in SPR cells the 
rates of transport to the reaction surface can be quite slow relative to the 
reaction rates, i.e., it may take much longer to diffusively transport down 
to the receptor surface than it does to bind to that surface, so the 
well-mixed assumption of first-order reaction kinetics may not necessarily 
be valid. The rate of transport to the reaction surface combines with the 
intrinsic reaction rates to create the effective reaction rates that are 
measured in an SPR assay. In order to determine the intrinsic reaction rates 
the influence of the transport rate must be properly accounted for 
\cite{Tauber}.
  
Most of the attempted approaches to the problem of decoupling
the transport and reaction rates model the system
 with a deterministic process, where the dependence on the 
parameters of the SPR system is governed by a set of coupled 
differential partial rate equations \cite{RT-Model, My1, My2}. Simulations 
for SPR systems are often derived from numerical solutions to 
these PDEs, but these solutions often fail to capture the spatial 
and temporal correlations between the ligands and the receptors 
as they interact, and ignore statistical fluctuations \cite{PDE-Sim}. 

Monte Carlo simulations are a computational tool developed to numerically solve the
basic master equation for stochastic processes, and faithfully encode account for
the presence of fluctuations and correlations in the modeled system. Monte Carlo
methods have found widespread application in the modeling of physical, chemical,
and biological systems. Since we cannot provide a comprehensive overview of Monte 
Carlo techniques in this brief paper, we refer the reader to Ref.~\cite{Monte_Carlo}
as a recent review of stochastic modeling for biological systems.

In this paper, we present results from Monte Carlo simulations of SPR cells for a broad 
range of binding and unbinding rates that allow for the observation of how the presence 
of correlations and fluctuations influence SPR data. Reaction rates derived from the 
standard mean-field model of the reaction kinetics\footnote{The mean-field model of 
reaction kinetics is physics nomenclature for the well-mixed assumption of the law of 
mass action (i.e., physical and temporal correlations are ignored). The term 
`mean-field' will be used to refer to this model throughout this paper, but the two 
terms are equivalent \cite{Tauber2, Rate-Equation}.} will be compared with known 
intrinsic reaction rates used in the simulations in order to determine the degree to 
which spatio-temporal correlations and fluctuations are important to the dynamics of 
the system.

\section{Surface plasmon resonance}
\subsection{The structure of the surface plasmon resonance cell}
The structure of the surface plasmon resonance cell is discussed
in more detail in literature \cite{RT-Model, Schasfoort}, but the following 
section will attempt to give a brief overview.

\begin{figure}
\centering
\includegraphics[scale=.4]{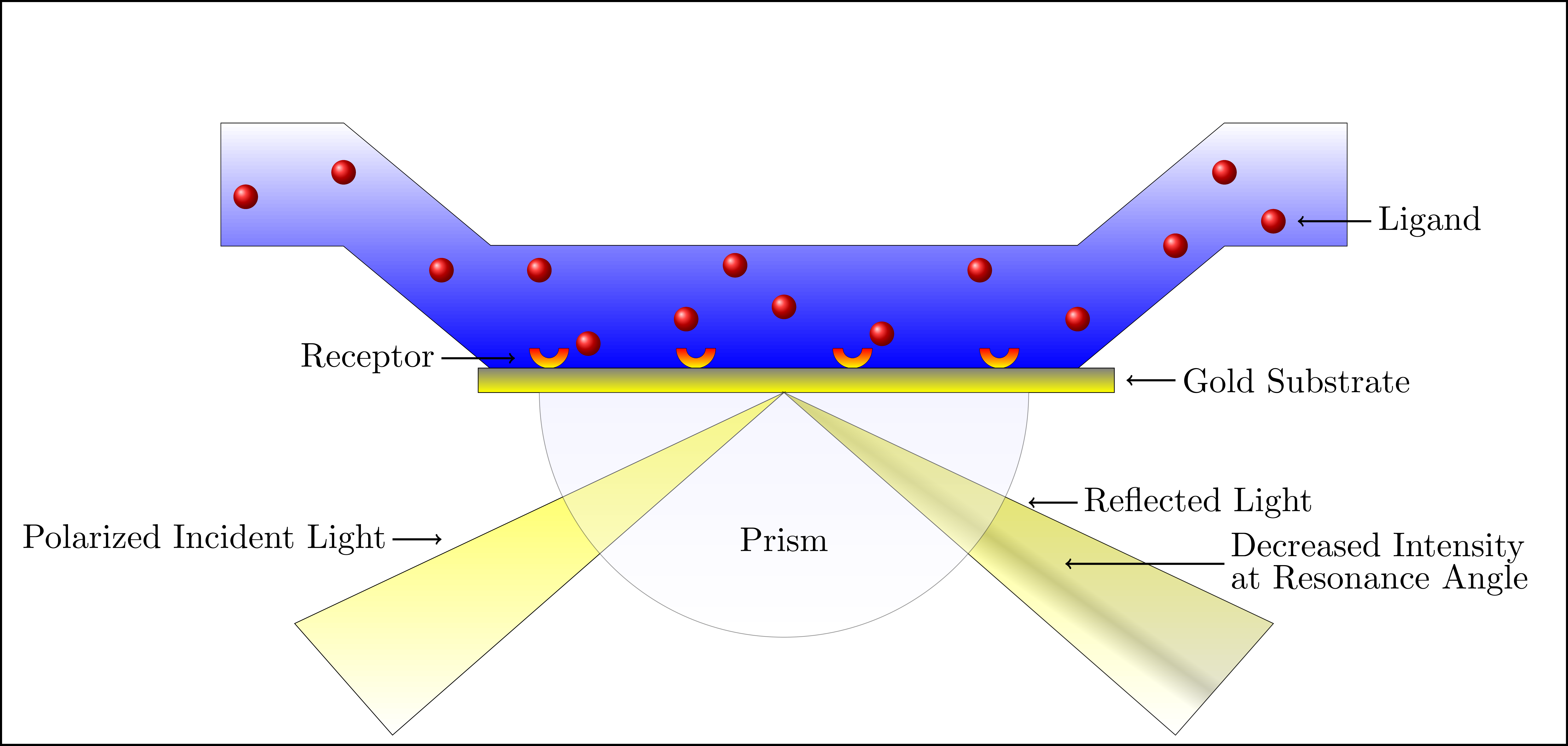}
\caption{Surface plasmon resonance cell (schematic). A gold substrate is embedded at the 
bottom of a flow cell. Receptors (red-orange half circles) are distributed evenly across 
the substrate. Ligands (red spheres) are dissolved in a non-reactive solvent, and allowed 
to flow across the receptor surface with a constant concentration and flow rate. Ligands 
will be transported down to the receptor surface where they bind and unbind to the 
receptors according to their dynamics. Incident p-polarized light is shown through a prism 
onto the receptor surface. The angle at which resonance between the incident beam and the 
standing waves of electrons (plasmons) in the gold substrate can be measured by recording 
the angle at which the reflected light has a decreased intensity \cite{SPR-Book, RT-Model, 
SPR_Procedure}. The extracted data of this resonance angle as a function of time can be 
rescaled to indicate the bound ligand density (i.e. the number of bound ligands normalized 
by the concentration of ligands in the flow cell) as a function of time \cite{RU-scaling}.
\label{fig:spr_cell}}
\end{figure}

A surface plasmon resonance cell (schematically detailed in Fig. \ref{fig:spr_cell}) is 
constructed by embedding a gold substrate into the bottom a of flow cell with linear 
dimensions on the order of millimeters. Two chemical species are chosen with the goal of 
determining the binding dynamics between them. One of these species is designated the 
receptors, and the other the ligands. The receptors are typically distributed randomly 
along the gold substrate and fixed in place, creating the receptor surface. A non-reactive 
solvent has a predetermined concentration of ligands dissolved into it, and this solution 
is allowed to flow over the receptor surface at a constant flow velocity.

The ligands in the solution are transported diffusely down to the receptor surface, where 
they bind and unbind to the receptors according to their respective dynamics. The binding 
and unbinding of the ligands to the receptors cause the resonance energy of the surface 
plasmon waves in the gold substrate to change \cite{SPR-Book, RT-Model, SPR_Procedure}. 
This change in the energy of the waves can be measured by shining a p-polarized beam of 
light onto the substrate through a prism. The prism allows the momentum of the incident
beam to be varied, and when the momentum of the incident beam and the surface plasmons of
the gold substrate are the same, the beam and plasmons couple and create a surface plasmon 
polariton in the gold substrate. This coupling results in a decrease in energy of the 
reflected beam of light, and the momentum at which this occurs can be measured by 
recording the angle where the resonance between the incident beam and the surface plasmon 
appears. The change in resonance angle as a function of time can then be rescaled into a 
plot of bound ligand-receptor pairs as a function of time \cite{RU-scaling}.

\subsection{Stages of the surface plasmon resonance experiment}
\begin{figure}
\centering
\includegraphics[scale=.75]{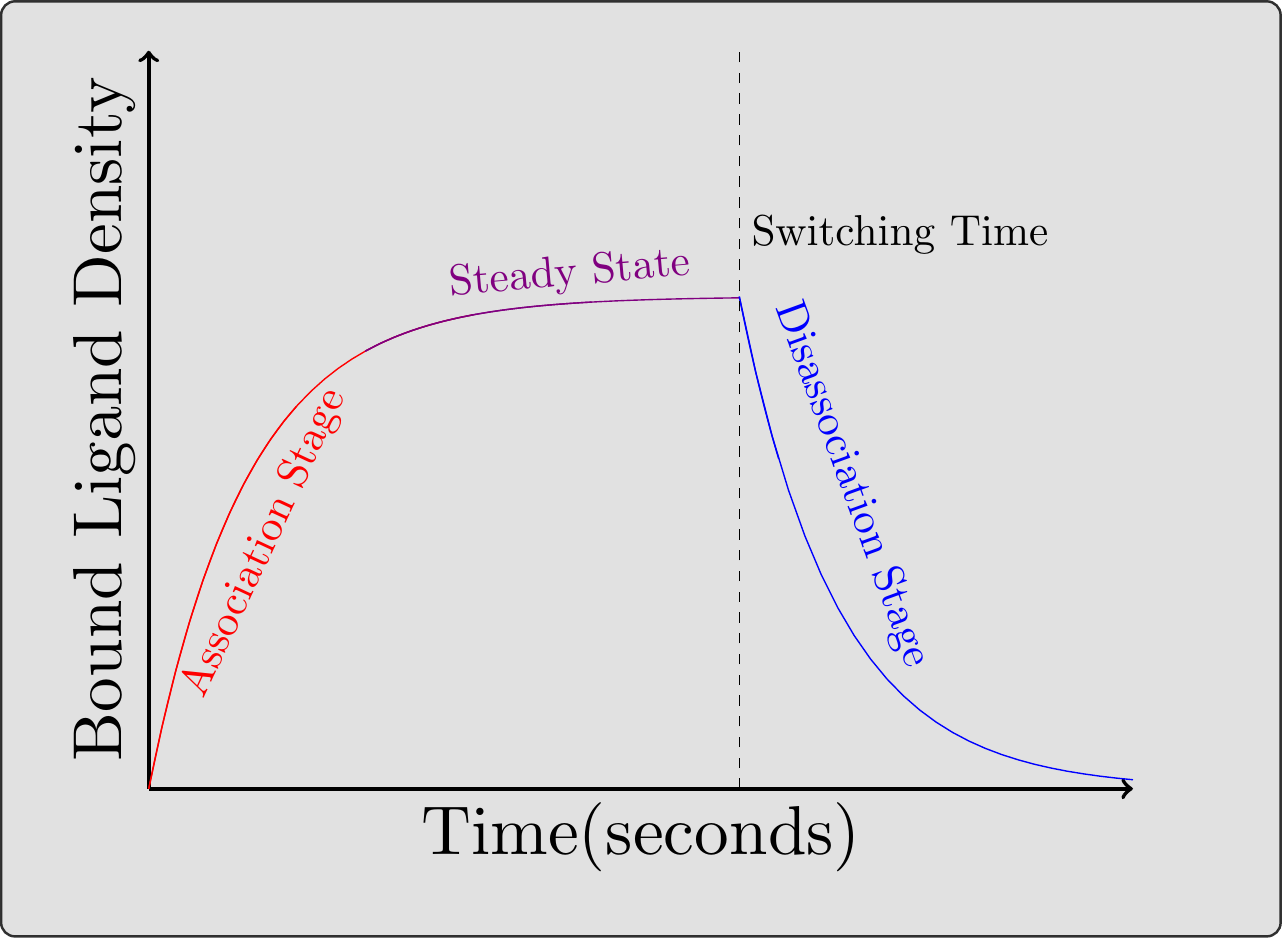}
\caption{Stages of an SPR experiment. The red line indicates the association state of the 
SPR experiment, where a solution of ligands flows over the receptor surface with a constant 
concentration and fixed flow velocity. The bound concentration reaches a steady state 
(depicted in purple), at which point the concentration of incoming ligands is cut off, 
letting the bound ligands decay off the receptors in the dissociation stage, represented by
the blue line.\label{SPR-Stages}}
\end{figure} 
The experimental process of surface plasmon resonance is typically performed in two stages.
First, the solution of ligands is allowed to flow over the receptor surface with a constant 
concentration of ligands and fixed flow velocity.  The system is allowed to evolve in 
this state until a steady-state concentration of bound ligands is observed. This stage of 
the experiment is referred to as the association stage. Subsequently, the concentration 
of incoming ligands is cut off, and the number of bound ligand-receptor pairs is allowed to 
decay away, as the ligands gradually unbind. This stage of the experiment is known as the 
dissociation stage. The concentration of bound ligands is measured throughout both stages, 
and data similar to the kind depicted in Fig.~\ref{SPR-Stages} is generated. This paper aims
to replicate both stages via Monte Carlo simulations, in order to determine the role that 
the spatio-temporal correlations induced by diffusion-limited association and repeated 
ligand rebinding processes play in the dynamics of the SPR cell.

\section{SPR cell model}

\subsection{Cell geometry}
We model the SPR cell as a rectangular lattice, with 
lattice spacing of 10nm. The lattice is constructed with maximum dimensions of $L_x,L_y,L_z$ 
on the $x$, $y$, and $z$ axes, which correspond to the laboratory dimensions of the SPR 
chip. Periodic boundary conditions are imposed on the $z$ axis, and a reflective boundary 
condition imposed along the $y=L_{y}$ top of the $y$ axis. Ligands are introduced 
at the $x=0$ surface, and perform a random walk to adjacent lattice sites until 
they encounter the $x=L_x$ surface, at which point they are removed from the lattice. 
\begin{figure}[h]
\centering
\includegraphics[scale=.75]{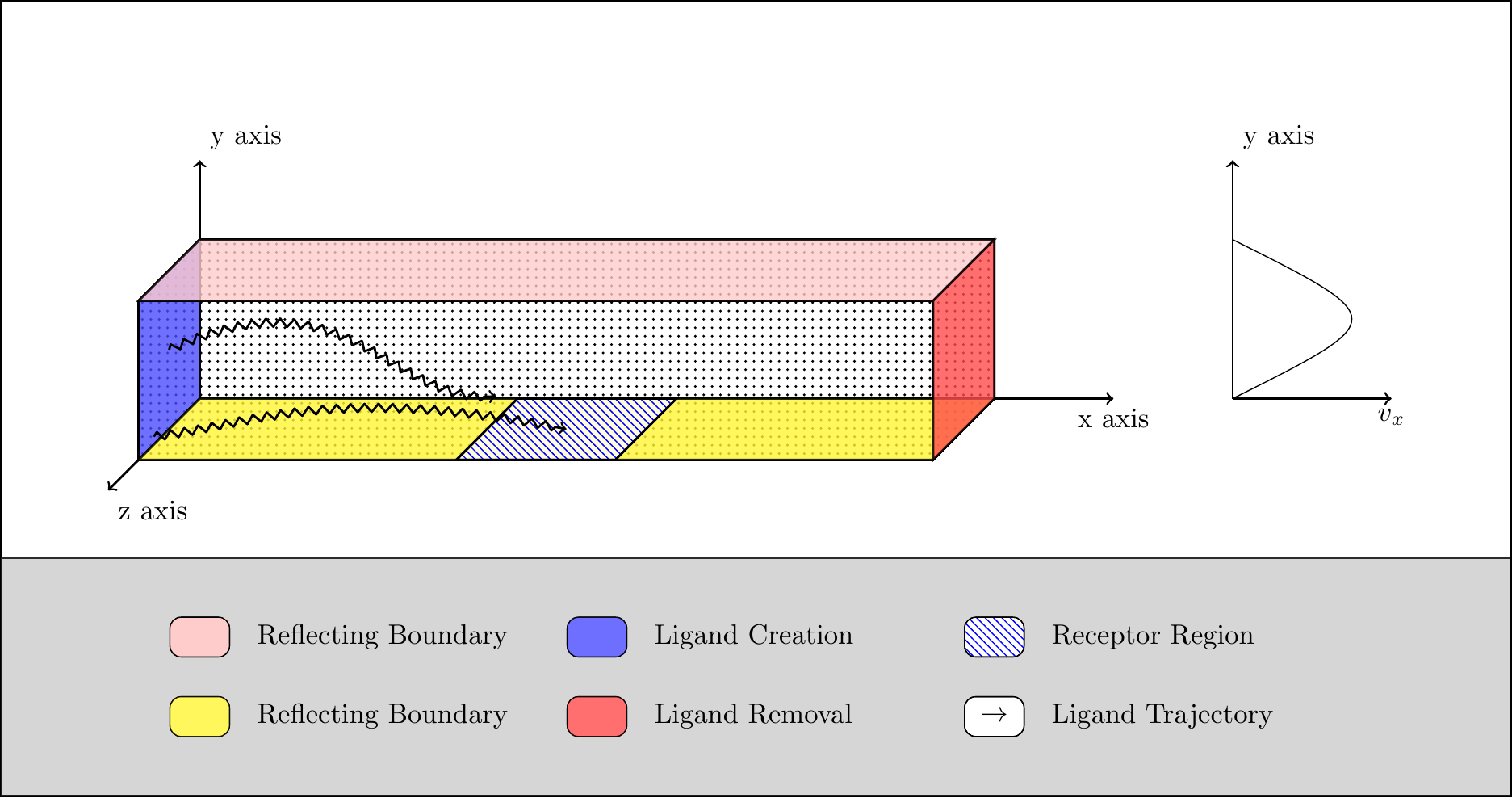}
\caption{The discretized model of the SPR cell. Ligands are introduced at the ligand 
creation region depicted in solid blue, and perform a random walk through the lattice. This 
random walk is biased to create a parabolic flow profile (shown in the plot to the right of 
the schematic) that would be expected in the regime of laminar flow typical of SPR cells. 
The top and bottom planes of the lattice (depicted by the solid pink and solid yellow planes)
form reflecting boundaries for the ligands. Receptors are evenly distributed in the receptor 
region (the dashed blue plane), and any ligand directly adjacent above a receptor has a 
chance to bind to it according to the association rate $\widetilde{k_{+}}$. Ligands perform 
their random walk through the lattice sites until they encounter the ligand removal region 
(depicted by the solid red plane), where they are removed from the simulation. In the 
association stage of the simulation, a ligand is immediately introduced at the ligand 
creation region to keep the concentration of ligands in the SPR cell constant, while in the 
dissociation stage of the simulation, the ligands are removed and not reintroduced, to allow
the concentration of bound ligands to decay. \label{fig:cell_model}}
\end{figure}

A subsection of the $y=0$ surface is selected to model the receptor surface, from $x=x_0$ to 
$x=x_1$. Receptors are distributed evenly over this subsection with density $R_0$, and the 
receptors are modeled such that if a ligand is directly adjacent above the receptor, the 
ligand can bind to the receptor with a probability $\widetilde{k_{+}}$. Once the ligand is 
attached to a receptor it can no longer move, but can unbind from the receptor with 
probability $\widetilde{k_{-}}$. Ligands are assumed to be small enough that they do not 
interact in the lattice, and a receptor that is bound to a ligand cannot bind to another 
ligand until the first ligand unbinds. 

A summary of the laboratory parameters of the SPR chip is given in Table~\ref{tab:SPR_Lab}, 
and a schematic representation of the simulation cell shown in Fig.~\ref{fig:cell_model}.

\begin{table}[h]
\caption{\label{tab:SPR_Lab}The laboratory parameters of a surface plasmon resonance chip.}
\begin{indented}
\item[]\begin{tabular}{@{}llll}
\br
Parameter &  Description & \multicolumn{2}{l}{Value} \\
\mr
$L_{x}$ & Lattice size along $x$ axis & 4.80  & $mm$  \\
$L_{y}$ & Lattice size along $y$ axis & 0.0500 & $mm$  \\
$v$ & Mean flow velocity & 1.33 & $mm/s$ \\
$D$ & Diffusion coefficient & 30.0 & $\mu m^2/s$  \\ 
$R_{0}$ & Receptor concentration & 5000 & $\mu m^{-2}$ \\ 
$C_{0}$ & Ligand concentration & 100 & $nM$ \\ 
$k_{+}$ & Association rate & --- & $M^{-1}s^{-1}$ \\ 
$k_{-}$ & Dissociation rate & --- & $s^{-1}$ \\ 
$x_{0}$ & Start of SPR scanning region & 2.9 & $mm$ \\ 
$x_{1}$ & End of SPR scanning region & 4.3 & $mm$ \\ 
\br
\end{tabular}
\end{indented}
\end{table}

While SPR regions are three-dimensional, the dynamics themselves are captured 
sufficiently in a two-dimensional representation if enough simulations are performed.
Thus, only the $x$ and $y$ dimensions of the SPR chip are of concern for the model.
The laboratory parameters are then discretized using the lattice constants detailed in 
Table~\ref{tab:Lattice_Constants}, which give the SPR model parameters listed in 
Table~\ref{tab:Model_Params}.
\begin{table}[h]
\caption{\label{tab:Lattice_Constants}The lattice constants used to discretize the SPR model.}
\begin{indented}
\item[]\begin{tabular}{@{}llll}
\br
Parameter &  Description & \multicolumn{2}{l}{Value} \\
\mr
$\lambda$ & Lattice size constant & 10 & $nm$ \\
$\delta t$ & Time step  & $1.51 \times 10^{-6}$ & {\it s}\\ 
\br
\end{tabular}
\end{indented}
\end{table}


\begin{table}[h]
\caption{\label{tab:Model_Params}The discretized parameters of the surface plasmon resonance 
model.}
\begin{indented}
\item[]\begin{tabular}{@{}lll}
\br
Parameter &  Relation to lab param. & Value \\
\mr
$\widetilde{L_{x}}$ & $L_{x}/\lambda$ & $4.80\times 10^{5}$  \\ 
$\widetilde{L_{y}}$ & $L_{y}/\lambda$ & $5\times 10^{3}$  \\ 
$\widetilde{v}$ & $v\cdot (\delta t/\lambda)$ & 200.8  \\ 
$\widetilde{D}$ & $D\cdot (\delta t/\lambda^{2})$ & 0.453  \\ 
$\widetilde{R_{0}}$ & $R_{0}\cdot \lambda^2$ & 0.5 \\ 
$\widetilde{C_{0}}$ & $C_{0}\cdot N_{A}\cdot \lambda^{3}$ & $6.022\times 10^{-5}$ \\ 
$\widetilde{k_{+}}$ & $k_{+} \cdot \delta t/(N_{A} \cdot \lambda^{3})$ & --- \\ 
$\widetilde{k_{-}}$ & $k_{-}\cdot \delta t$ & --- \\ 
$\widetilde{x_{0}}$ & $x_{0}/\lambda$  & $2.90\times 10^{5}$ \\ 
$\widetilde{x_{1}}$ & $x_{1}/\lambda$ & $4.30 \times 10^{5}$ \\ 
\br
\end{tabular}
\end{indented}
\end{table}

\subsection{Ligand movement}
Surface plasmon resonance cells are small, on the order of millimeters. 
This results in SPR cells having very small Reynolds numbers \cite{reynolds_number}. This 
in turn means that SPR cells reside in the regime of almost ideal laminar flow, so the
movement of ligands in our simulation is biased to reflect this laminar transport. 
\begin{figure}[h]
\centering
\includegraphics[scale=.3]{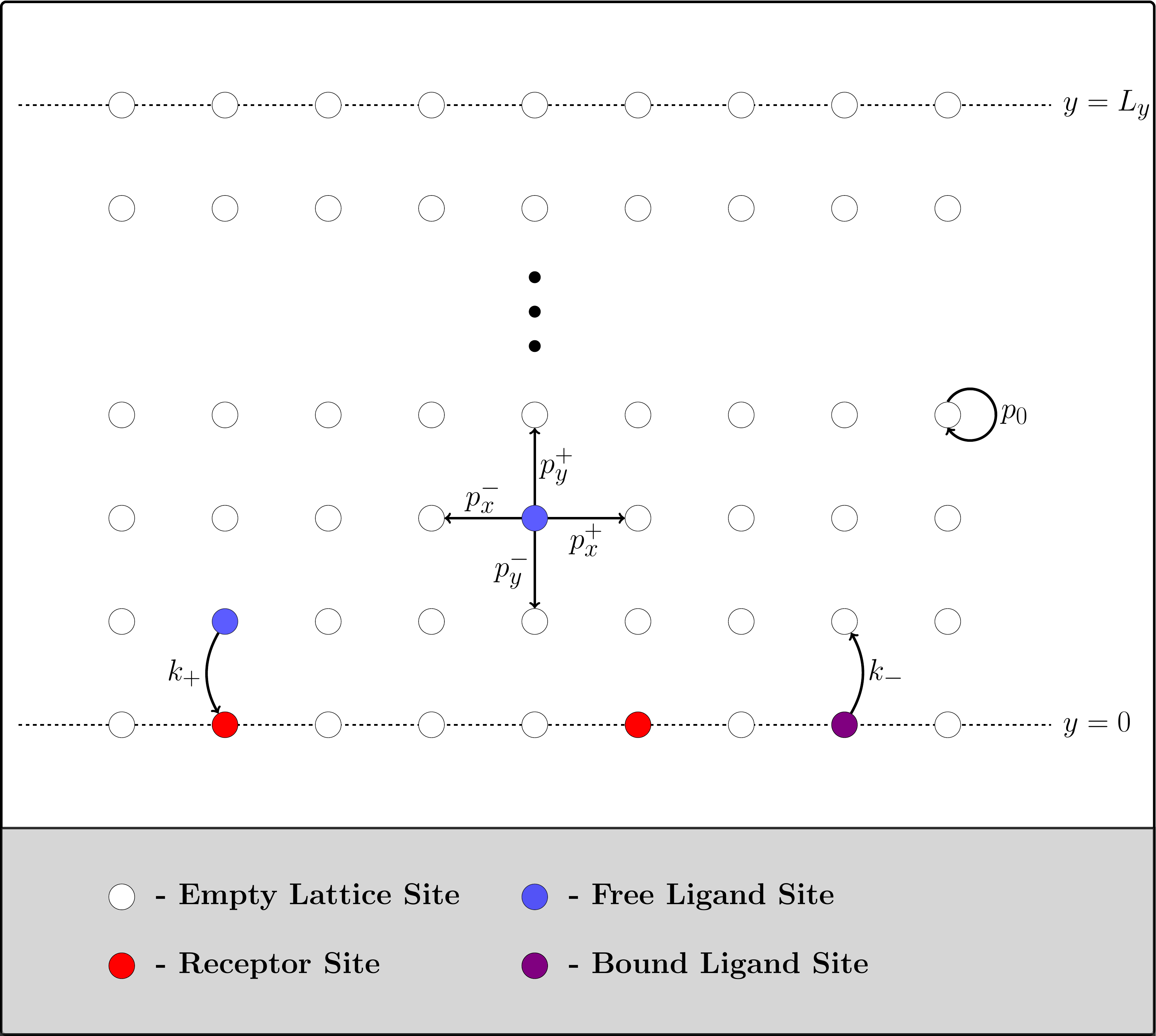}
\caption{The two-dimensional dynamics of the SPR model. This plot shows all the possible 
actions that a ligand might take as it moves through the lattice sites. A ligand has 
probabilities $p_{\mu}^{\pm}$ of moving forward (+) or backwards (-) in the 
$\mu \in \{x,y,z\}$ direction, and a probability $p_{0}$ of stay put. Additionally, if a 
ligand is above a receptor it has probability $\widetilde{k_{+}}$ of binding to the receptor
(independent of the probabilities of movement; for the purposes of the simulation the 
ligands are stepped in a direction determined by the movement probabilities, and then check 
if they could bind to a receptor). If a ligand is bound, it can no longer move, but has a 
probability $\widetilde{k_{-}}$ to unbind. Once unbound, the ligand continues the random 
walk through the lattice.\label{fig:move_model}}
\end{figure}

The movement of the ligands through the lattice is  modeled via a biased random walk, 
where the probabilities of moving parallel to the flow velocity are adjusted to create 
a parabolic flow profile as is expected in the case of laminar flow. The first moment 
of the ligand position is taken from the flow velocity in that direction, 
\begin{equation}
p_{\mu}^{+} - p_{\mu}^{-} = \widetilde{v}_{\mu} \, ,
\end{equation}
The second cumulant of the ligand position is taken from diffusion in the fluid,
\begin{equation}
 (p_{\mu}^{+} + p_{\mu}^{-}) - (p_{\mu}^{+} - p_{\mu}^{-})^{2} = \widetilde{D_{\mu}} =
 \widetilde{D}/3 \, .
\end{equation}
The probabilities of ligand movement can be extracted from these conditions along with a 
normalization condition:
\begin{equation}
p_{0} + \sum_{\mu} p_{\mu}^{\pm}=1 \, .
\end{equation}
Here $p_{0}$ is the probability of staying still, $p_{\mu}^{\pm}$ respectively denote the 
probability of moving in the positive or negative $\mu$ direction; $v_{\mu}$ and $D_{\mu}$ 
are the flow velocity and diffusion constant in the $\mu$ direction, where $\mu$ can be 
either $x$, $y$, or $z$. Diffusion in the system is isotropic while the following bias 
velocities are chosen to model laminar flow:
\begin{eqnarray}
&&\widetilde{v_{y}}=\widetilde{v_{z}}=0 \, , \\
&&\widetilde{v_{x}}=\frac{6\widetilde{v}y(\widetilde{L_{y}}-y)}{\widetilde{L_{y}}^2} \, .
\end{eqnarray}
The probabilities of movement perpendicular to the flow velocity are unchanged. The 
parameters with a `$\sim$' superscript are dimensionless simulation parameters related to 
the physical parameters of the SPR chip via Table~\ref{tab:Model_Params}. The dimensional 
mean flow velocity $v$ is related to the pressure gradient $\Delta P$ across the system as 
well as the viscosity $\eta$ \cite{LL-Fluid} via
\begin{equation}
v=-\frac{L_y^{2}\Delta P}{12\eta L_{x}} \, .
\end{equation}

As the ligands propagate through the lattice and encounter receptors in the receptor surface 
on the lattice floor, some percentage of the ligand population will bind to the receptors. 
This percentage is measured every time step for both the association and dissociation stages of the simulation. 
An example of these results is shown in Figure~\ref{fig:example}. A brief summary of the algorithm used for the Monte Carlo simulations is given in Appendix C.

\subsection{Analysis}
\begin{figure}
\centering
\includegraphics[width=0.75\textwidth]{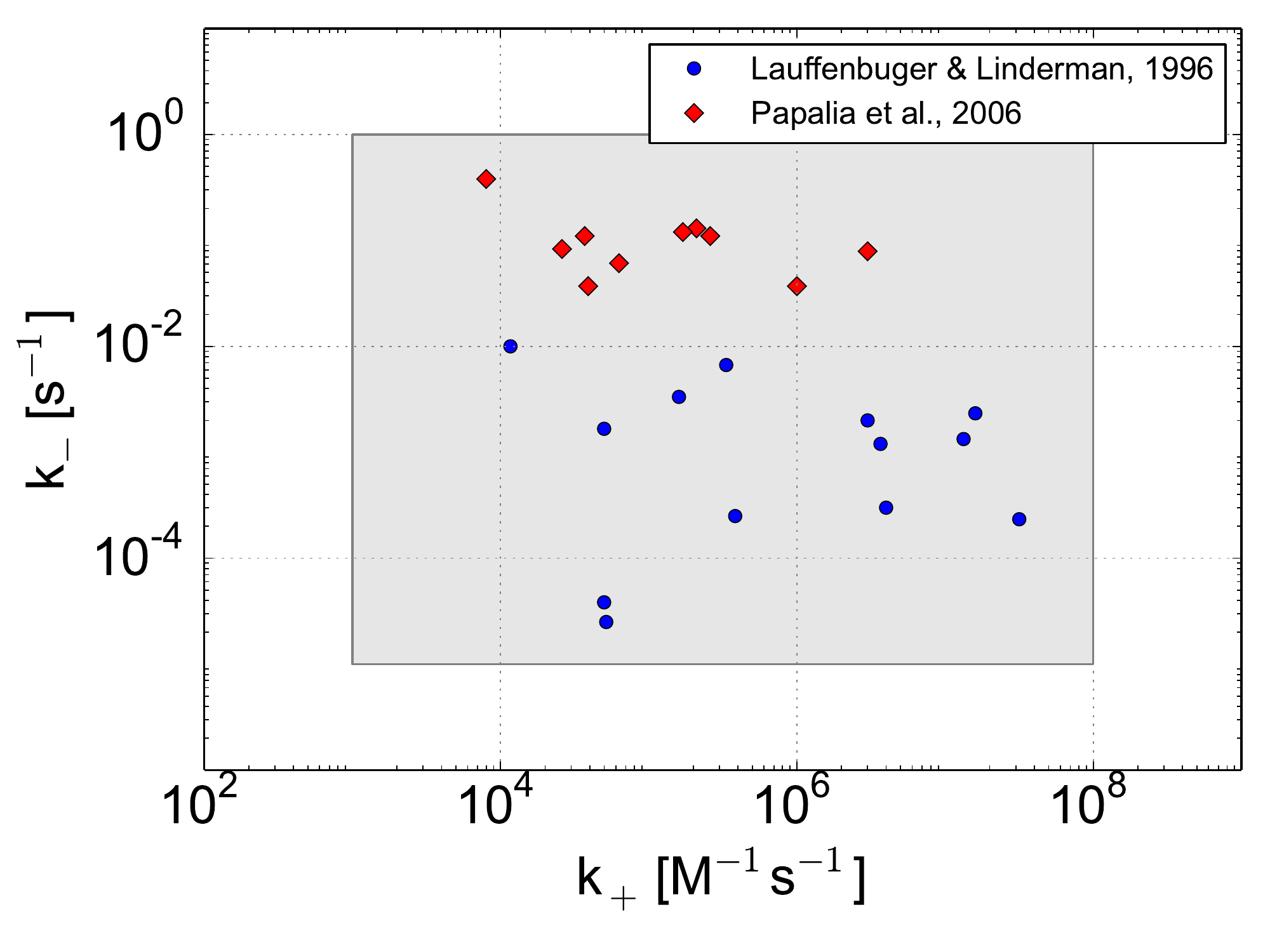} \vskip -0.5cm
\caption{The range of experimentally determined reaction rates between pairs of different 
chemical species. The dissociation rate $k_{-}$ of the chemical pair is plotted against the 
association rate $k_{+}$ on a log-log plot in order to give a representation of the range 
of values that these rates can take. The blue circles represent pairs recorded by Papalia 
et al. \cite{Papalia-Rates}, while the red diamonds represent pairs recorded by 
Lauffenburger and Linderman \cite{LL-Rates}. The shaded region represents the regime of 
typical association and dissociation rates.\label{fig:ranges}}
\end{figure}

The system described in Table~\ref{tab:Model_Params} was then simulated, with the parameters 
scaled by a factor of $\alpha=0.025$ as described in Appendix B. Nine different association rates and two different 
dissociation rates were selected from the range of known values (detailed in 
Fig.~\ref{fig:ranges}, with values ranging from $10^{3}M^{-1}s^{-1}$ to $10^{7}M^{-1}s^{-1}$ 
and $10^{-2}s^{-1}$ to $10^{-3}s^{-1}$ respectively). 

\begin{figure}[h]
\centering
\includegraphics[scale=.6]{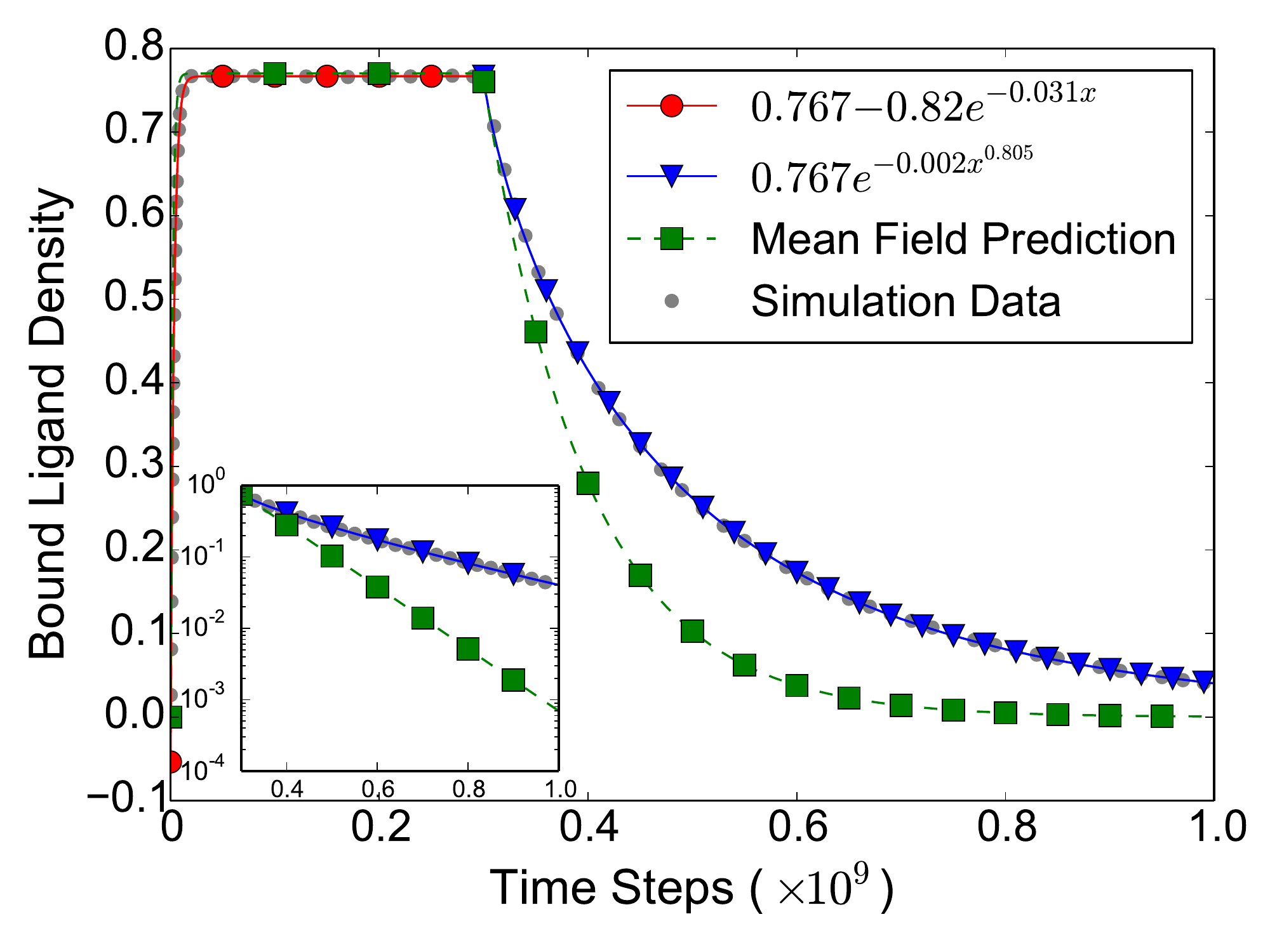} \vskip -0.5cm
\caption{An example of simulation data for an association rate of $10^{6}M^{-1}s^{-1}$ and 
a dissociation rate of $10^{-3}s^{-1}$. Error bars are the same size as the data points, 
and are thus excluded. The simulation results for the density of bound ligands is 
represented by the (gray) dots. A subset of the simulation data points is shown to 
ensure that the data points do not overlap and are easily visible. The fit of the 
association stage of the simulation is 
represented by the (red) line with circular markers, and the fit of the dissociation stage 
is represented by the (blue) line with triangular markers. For comparison, the mean-field 
prediction for a dissociation rate of $10^{-3}s^{-1}$ is shown by the (green) squares.  
The inset is a logarithmic (base ten) plot of the dissociation data of the main 
panel, again plotting the bound ligand density versus simulation time steps. The (blue) 
line marked with triangles is the stretched exponential fit of the data, represented by the
(gray) dots, and the mean field prediction is represent by the (green) squares. This 
particular rate pair was selected because it demonstrates the non-exponential behavior of 
the dissociation phase at high $Da$. This is easily seen in the form of the fit for the 
dissociation phase, which is a stretched exponential (i.e., 
$p(t)\propto e^{-\alpha t^{\beta}}$ for $\alpha,\beta \in \mathbb{R}$) rather than simple 
exponential (i.e., $p(t)\propto e^{-\alpha t}$ for $\alpha \in \mathbb{R}$). This 
contradicts the predictions of the mean-field analysis, and will be discussed in more 
detail in Sec.~4. \label{fig:example}}
\end{figure}

All possible pairs of these association and dissociation rates where then simulated giving 
eighteen different simulations. In order to obtain statistically significant results, each 
of these eighteen simulations was performed five hundred times (each time the simulation 
is independent of all others), with new random initial conditions for each realization of 
the simulation. The number of realizations of each simulation was chosen to be five hundred
in order to shrink the associated error while still being computationally feasible. 
Figure~\ref{fig:example} shows example results of an averaged set of five hundred runs of 
an association-dissociation rate pair simulation. The example simulation data in 
Fig.~\ref{fig:example} displays fits for both the association stage (red circles), and the 
dissociation stage (blue triangles). The mean field prediction of the dissociation phase 
is represented by the (green) dashed line with square markers.  The error bars are not 
included because they are the same size as the (gray) data points. The inset in 
Figure~\ref{fig:example} highlights the non-exponential behavior of the dissociation phase,
by showing a logarithmic plot of the dissociation stage of Fig.~\ref{fig:example}. 
The (blue) line with triangular markers is the non-exponential fit of the (gray) data 
points, and the (green) dashed line with square markers is the mean-field prediction. 
Again, error bars are excluded because they are the same size as the (gray) data points. 
This plot of a high association rate is chosen to showcase the non-exponential behavior of the dissociation stage at high $Da$. This behavior does not 
coincide with the prediction of the mean-field analysis, and will be discussed in Section~4.

\subsection{Mean-field approximation}

The mean-field rate equation for the SPR system is given by the first-order differential 
equation for the bound ligand concentration $p$\footnote{In this case $p$ is defined as the number 
of bound ligand-receptor pairs normalized by the number of ligands in the volume of the SPR cell 
bounded by the receptor surface. This number of ligands has a value of: $n_{l}=C_{0}(x_{1} - x_{0})L_{y}L_{z}$.},
\begin{equation}
\label{eq:mf}
\dot{p}=C_{0}k_{+}(\gamma-p)-k_{-}p \, ,
\end{equation}
Where $C_{0}$, $k_{+}$, and $k_{-}$ are described in Table \ref{tab:SPR_Lab} and 
$n_{l} = C_{0}(x_{1}-x_{0})L_{y}L_{z}$ and $n_{r} = R_{0}(x_{1}-x_{0})L_{z}$ are the number of 
ligands and receptors in the SPR scanning region, respectively. The factor $\gamma$ is the 
ratio of the number of ligands in the volume of the SPR cell bounded by the receptor surface, to the number of receptors on the receptor surface: $\gamma = n_{l}/n_{r}$.

The mean-field association and dissociation rates were extracted via several parameters 
(summarized in Table~\ref{tab:metric}) that are easily extracted from the numerical data.
These values are often employed in the analysis of 
sensogram\footnote{A sensogram is a plot of SPR data vs. time. Figure~\ref{fig:example} 
is an example sensogram, generated via simulations.} data \cite{Glaser, Schuck-Minton}.
The mean-field model, eq.~(\ref{eq:mf}), provides predictions for these parameters which 
are summarized in eqs.~(\ref{eq:f0})-(\ref{eq:r_finty}) below.  
Specifically, the parameters listed in Table~\ref{tab:metric} are: the time derivative 
$f_{0} = \dot{p}(0)$ of the bound ligand concentration at the initial time\footnote{Because
the concentration of ligands in the flow cell is not constant at the beginning of the 
simulation, the time used to calculate this was not $t=0$, but instead the time when the 
concentration began to behave like an exponential.}; $f_{\infty}$, which is the change in 
the time derivative $\dot{p}$ with respect to the bound ligand concentration $p$ at the 
switching time between the association and dissociation stages; the change $r_{0}$ in 
$\ln (p)$ with respect to time at the switching time; the change $r_{\infty}$ in $\ln (p)$ 
with respect to time as time goes to infinity; and the saturation concentration $p^{*}$ of
bound ligands as they reach a steady state in the association phase:
\begin{eqnarray}
& f_{0} =&\gamma k_{+}C_{0} \, , \label{eq:f0} \\
& f_{\infty} =& k_{+}C_{0}+k_{-} \, , \label{eq:finty} \\
& p^{*} =& \frac{\gamma k_{+}C_{0}}{k_{+}C_{0}+k_{-}} \, ,\\
& r_{0} =& k_{-} \, , \\
& r_{\infty} =& k_{-} \, . \label{eq:r_finty}
\end{eqnarray}

\begin{center}
\begin{table}[h]
\caption{\label{tab:metric}The sensogram metrics.}
\begin{indented}
\item[]\begin{tabular}{@{}ll}
\br
Parameter &  Definition\\
\mr
$f_{0}$ & $\dot{p}(0)$\\ 
$f_{\infty}$ & $-\lim_{p\rightarrow p^{*}}(\frac{\partial^{2}}{\partial p\partial t}p)$  \\
$r_{0}$ &  $-\frac{\partial}{\partial t} \ln p(t) |_{t=t_{\rm switch}}$  \\ 
$r_{\infty}$  & $-\frac{\partial}{\partial t} \ln p(t) |_{t=t_{\infty}}$  \\ 
$p^{*}$  & $p(t_{\rm switch})$\\
\br
\end{tabular}
\end{indented}
\end{table}
\end{center}

To measure the association and dissociation rates, $f_{0}$, $f_{\infty}$, and $r_{0}$ were 
used. These parameters were chosen because they are easily extracted from the numerical 
data, and provide simple relations to the association and dissociation rates. The numerical 
values of each of the three parameters was taken from the simulation data for each of the 
rate pairs, and the association rates and dissociation rates were solved for twice, namely 
via
\begin{equation}
\label{eq:kp0}
k_{+} = \frac{f_{0}}{\gamma C_{0}} \, , 
\end{equation} 
or
\begin{equation}
\label{eq:kpinfty}
k_{+} = \frac{f_{\infty}-r_{0}}{C_{0}} \, .  
\end{equation}
In each case the dissociation rate of the system is
\begin{equation}
\label{eq:km}
k_{-}=r_{0} \, .
\end{equation}
The two different association rates $k_{+}$ are paired with the one dissociation rate 
$k_{-}$, and compared with the actual input simulation values of these rates.


\section{Results}

\begin{figure}[h]
\centering
\includegraphics[scale=.7]{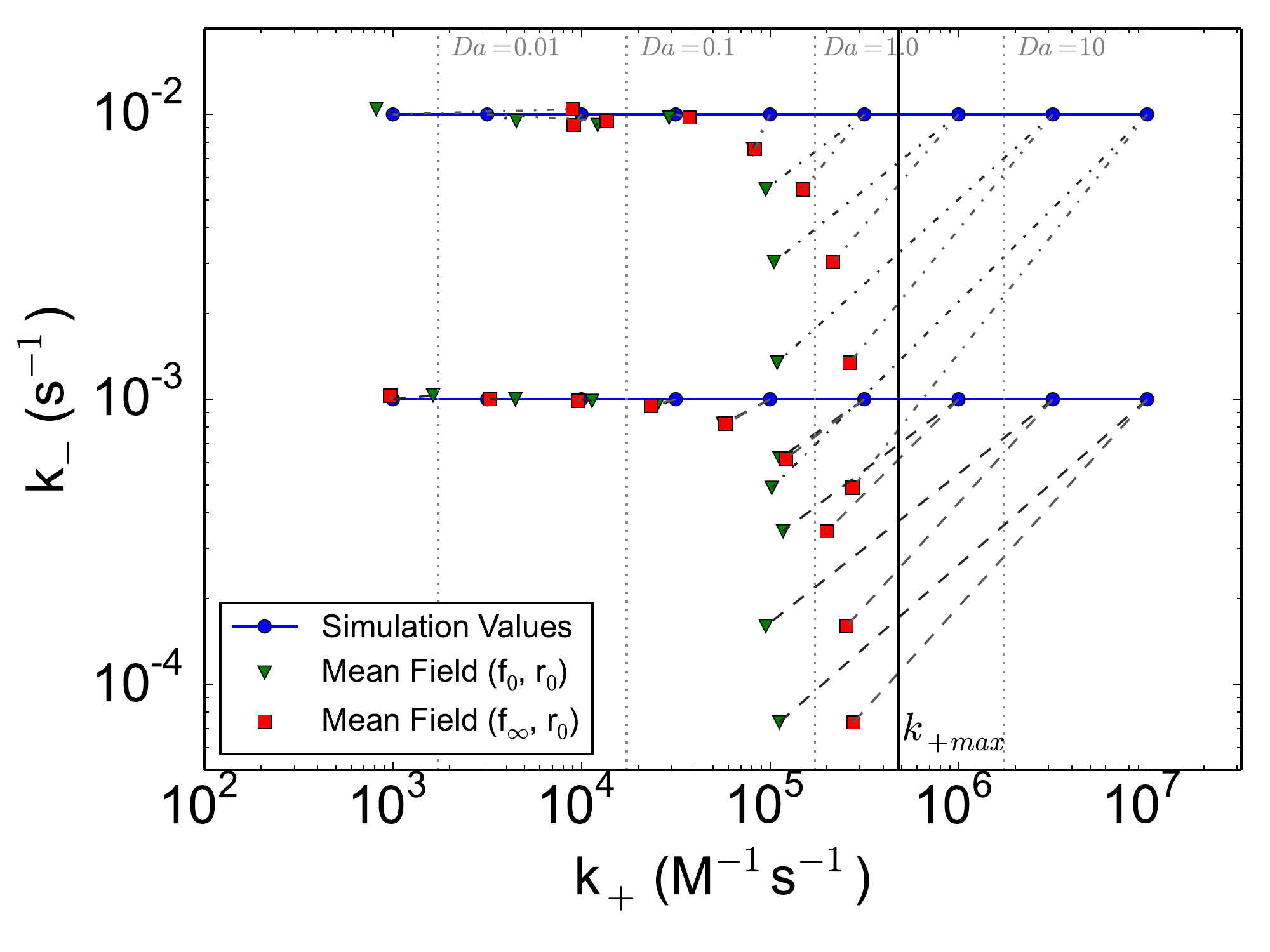} \vskip -0.5cm
\caption{The comparison of extracted and simulation association and dissociation rates. The 
plot shows the dissociation rates $k_{-}$ plotted against the association rates $k_{+}$ on 
a log-log scale for the eighteen different simulated pairs of association and dissociation 
rates. The intrinsic simulation rates are denoted by the (blue) circles, the rates extracted 
using the $f_{0}$ and $r_{0}$ sensogram metrics, eqs.~(\ref{eq:kp0}) and (\ref{eq:km}), are 
denoted by the (green) triangles, and the rates extracted by $f_{\infty}$ and $r_{0}$ 
sensogram metrics, eqs.~(\ref{eq:kpinfty}) and (\ref{eq:km}), are indicated by the (red) 
squares. The (gray) dashed lines connect the mean-field rates with the corresponding 
simulations from which they were extracted from. The dotted lines denote different values of 
constant $Da=k_{+}R_{0}(L_{x}L_{y}/6vD^{2})^{1/3}$. The solid (black) line labeled $k_{+max}$ 
represents a theoretical maximum that can be extracted from the mean-field theory for this 
particular system. Note that the highest value of $k_{+}$ that can be accurately predicted 
is much lower, and occurs around $Da\sim 0.1$. 
\label{fig:res}}
\end{figure}

The comparison of the simulation rates and the rates extracted from the data by applying the 
mean-field analysis can be seen in Fig.~\ref{fig:res}. The true simulation rates are denoted 
by the (blue) circles, the rates extracted using $f_{0}$ and $r_{0}$, eqs.~(\ref{eq:kp0}) and
(\ref{eq:km}), are denoted by the (green) triangles, and the rates extracted by $f_{\infty}$ 
and $r_{0}$, eqs.~(\ref{eq:kpinfty}) and (\ref{eq:km}), are indicated by the (red) squares. 
The (gray) dashed lines connect the mean-field rates with the corresponding simulations that 
they were extracted from. The dotted lines denote different values of constant 
$Da=k_{+}R_{0}(L_{x}L_{y}/6vD^{2})^{1/3}$. The solid (black) line labeled $k_{+max}$ marks a 
theoretical maximum that the mean-field theory can predict, which will be discussed below. 
These results were replicated with various values of the lattice spacing constant $\lambda$ 
and time step $\Delta t$ in order to ensure these results are independent of the discretization 
of the system. The values used in this paper were chosen because they accurately model the average 
receptor size and binding timescale of a SPR cell.

It is immediately apparent from Fig.~\ref{fig:res} that the extracted mean-field rates 
diverge rapidly from the simulation values as $Da$ increases, though it is interesting to 
note that the mean-field measurements of $k_{+}$ using $f_{0}$ and $r_{0}$ are better than 
those using $f_{\infty}$ and $r_{0}$ for $Da<0.1$ and high $k_{-}$, while the the predictions
of $f_{\infty}$ and $r_{0}$ are slightly more accurate for $Da>0.1$ than those of $f_{0}$ and
$r_{0}$. The better predictive abilities of $(f_{0},r_{0})$ at low $Da$ and high $k_{-}$ are 
due to the high sensitivity of the association rate $k_{+}$ to the sensogram metric 
$f_{\infty}$ at low $Da$ and high $k_{-}$.

\subsection{Sensitivity}

\begin{figure}[h]
\centering
\includegraphics[width=.73\linewidth]{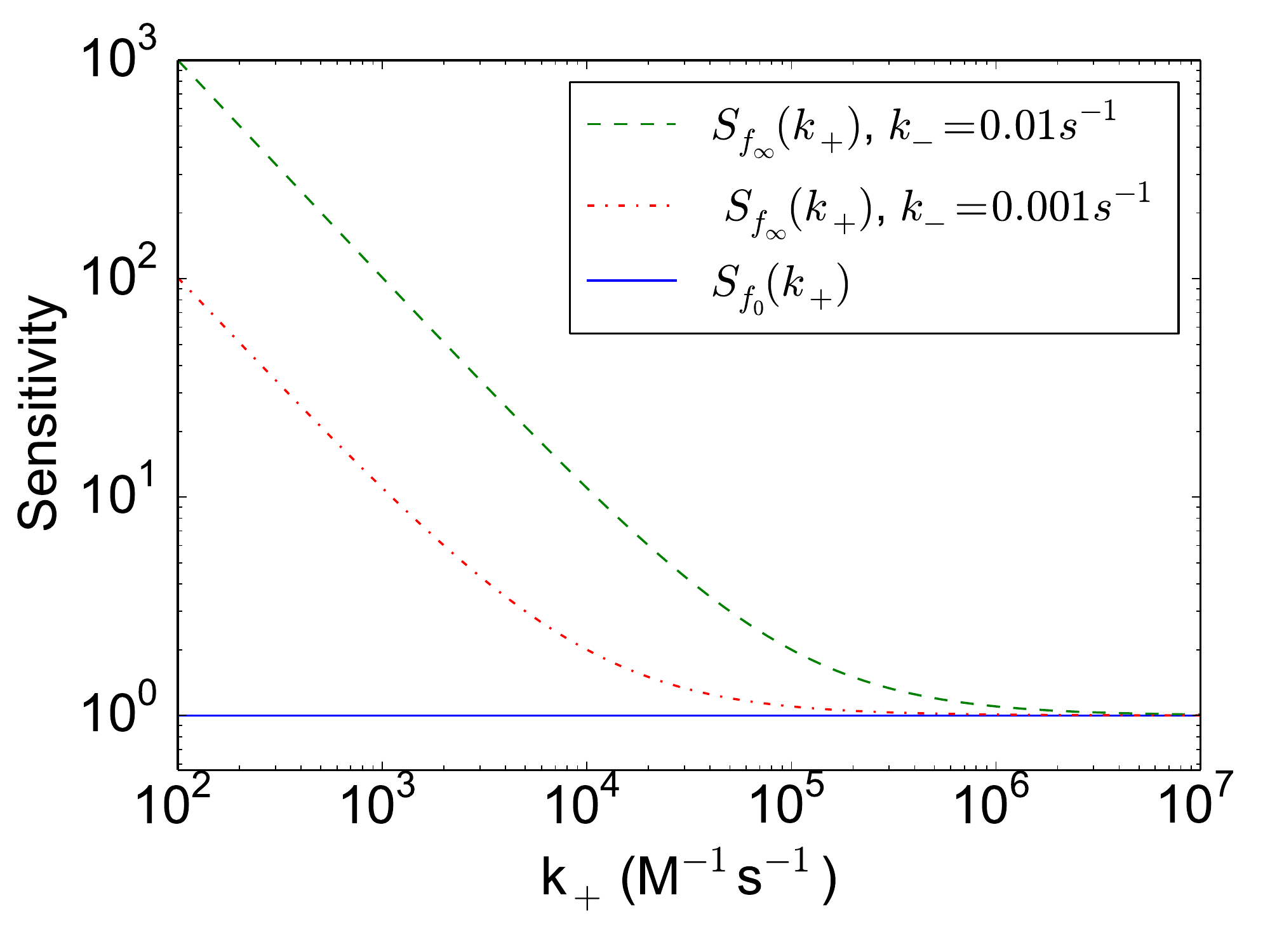} \vskip -0.5cm
\caption{A log-log plot of the sensitivity of the attachment rate to the sensogram metrics 
$f_{0}$ and $f_{\infty}$, eqs.~(\ref{eq:f0}) and (\ref{eq:finty}), as a function of $k_{+}$. 
The (green) dashed line is the sensitivity of $k_{+}$ to $f_{\infty}$ for $k_{-}=0.01s^{-1}$, 
the (red) dashed-dotted line is the sensitivity of $k_{+}$ to $f_{\infty}$ for 
$k_{-}=0.001s^{-1}$, and the (blue) solid line is the sensitivity of  $f_{0}$ to $k_{+}$ for 
all values of $k_{-}$. The concentration of ligands $C_{0}$ was taken to be 100nM.
\label{fig:sens}}
\end{figure}

Sensitivity in this context means the ratio of relative change in the extracted rate to the
relative change in the sensogram metrics. To clarify, if $y = f(x)$, then the sensitivity 
$S_{y}$, of $y$ to $x$ is defined by the relation $dy/y = S_{y}dx/x$. Thus 
$S_{y}(x)=(x/f(x))df/dx$. The sensitivity of $k_{+}$ to $f_{0}$ and $f_{\infty}$ is given by the equations
\begin{eqnarray}
&& S_{k_{+}}(f_{0})=1 \, , \\
&& S_{k_{+}}(f_{\infty}) = \frac{f_{\infty}}{f_{\infty}-r_{0}} 
   = \frac{C_{0}k_{+}+k_{-}}{C_{0}k_{+}}=1+K \, . 
\end{eqnarray}
For extraction of rate constants, the ideal value for sensitivity is $1$; sensitivities 
$\ll 1$ would indicate that the rate constants are independent of the sensogram metrics, 
while sensitivities $\gg 1$ indicate that small errors in the measurement of sensogram 
metrics will be amplified into large errors in the interpreted rate constants.
The sensitivities are plotted in Fig.~\ref{fig:sens} for the range of $k_{+}$ values used in the 
simulations, as well as both values of $k_{-}$. The (green) dashed line is the sensitivity 
of $k_{+}$ to $f_{\infty}$ with a constant $k_{-}=0.01s^{-1}$, the (red) dashed-dotted line
is the sensitivity of $k_{+}$ to $f_{\infty}$  with a constant $k_{-}=0.001s^{-1}$, and the 
solid (blue) line is the sensitivity of $k_{+}$ to $f_{0}$ for all values of $k_{-}$. As can 
be seen, in the regime where $k_{+}$ is relatively low and therefore $Da < 1$, $k_{+}$ is 
less sensitive to changes in the the sensogram metric $f_{0}$ than $f_{\infty}$. The results 
extracted from the $(f_{0},r_{0})$ interpretation therefore predict the rates more accurately
in this regime. Additionally, $k_{+}$ is approximately an order of magnitude less sensitive 
to $f_{\infty}$ for the smaller $k_{-}$ at low $Da$, and so the the predictions of the 
$(f_{\infty}, r_{0})$ metric at $k_{-}=0.001s^{-1}$ are more accurate than those of the same 
interpretation at $k_{-}=0.01s^{-1}$ for low $Da$.

\subsection{The diffusion-limited regime}

In the regime of high $Da$, $f_{\infty}$ becomes the more accurate of the the metrics. This 
(as noted in Ref.~\cite{Glaser}) is because $f_{\infty}$ is less affected by the transport 
of ligands, since it is extracted from later parts in the experiment, where most of the 
ligands in the system are near the binding surface. There is still a qualitative increase 
in the error of the sensogram metrics' predictions as $Da$ increases. One cause of this 
deviation is the effect of diffusive transport on the ligands. As $k_{+}$ increases, the 
average time for a ligand to bind to a receptor begins to be dominated by the time it takes 
for a ligand to be transported to the receptor surface \cite{Glaser}; however, at low 
association rates, $C_{0}k_{+}<D/(L_{y}/2)^2$, the time delay an average ligand will 
experience before binding will be due to the association rate. As the association rate 
increases into the regime of $C_{0}k_{+}>D/(L_{y}/2)^2$, the time delay will not be due to 
the association rate, but instead will be dominated by the much longer time it takes to be 
diffusely transported to the receptor. 

The mean-field approximation can only interpret the time spent before binding as being due to 
the association rate, and so the time scale it takes to diffusely transport ligands to the 
receptor surface gives a theoretical maximum on the association rate that the mean-field 
theory can predict, 
\begin{equation}
k_{+max}\approx \frac{1}{C_{0}}\frac{D}{(L_{y}/2)^{2}} \, .
\end{equation}
This value is marked with a (black) solid line in Fig.~\ref{fig:res}. In this figure the 
asymptotic approach of the $(f_{\infty},r_{0})$ prediction comes close to this value as $Da$
increases, while the prediction of $(f_{0},r_{0})$ approaches an asymptote at a lower value 
because it is more sensitive to the diffusive transport in the system.

\subsection{Ligand-receptor rebinding events}

The remaining effect to mention is that of ligand rebinding, which is assumed not to happen 
in the mean-field dissociation phase of the SPR experiment. However, the ligands may still 
perform random walks back to the receptor surface after they have unbound. As the association
rate increases, the likelihood of a ligand rebinding to a receptor increases. This causes 
ligands to on average stay on the receptor surface longer. The mean-field interpretation of 
this is a lowered dissociation rate, which is why the extracted dissociation rate decreases 
as the simulation association rate increases.

Additionally, it was predicted by Gopalakrishnan et al. \cite{Tauber} that ligand 
dissociation from a surface with uniform receptor density $R_{0}$ into a semi-infinite 
domain in the absence of advective transport results in non-exponential late time 
dissociation of the form $p(t)\propto e^{ct}{\rm erfc}(ct)$ where $c$ is a parameter that
depends on the density of receptors and the dissociation rate, and 
${\rm erfc}(z)=2/\sqrt{\pi}\int_{z}^{\infty}e^{-x^{2}}dx$. As seen in 
Fig.~\ref{fig:example}, the dynamics of the dissociation phase are indeed non-exponential 
for high $Da$, but are stretched exponentials (i.e. $p(t)\propto e^{-\alpha t^{\beta}}$ for 
$\alpha,\beta \in \mathbb{R}$) instead of error functions. This difference from the 
predictions of Ref.~\cite{Tauber} is likely due to the presence of advective transport in 
the SPR cell. For low $Da$, the behavior of the late-time dissociation corresponds to 
exponential kinetics, as the effects of the temporal correlations of ligand-receptor 
rebinding and diffusion are negligible compared to the time it takes for association. This 
exponential behavior at low $Da$ corresponds to the agreement between the simulation rates 
and the mean-field predictions at low $Da$, as seen in Fig.~\ref{fig:res}.

\section{Conclusion}

These Monte Carlo simulations of ligand-receptor binding kinetics in SPR cells provide a 
testing ground for different analysis techniques. They were used in this paper to determine 
the regime in which a mean-field analysis of SPR is applicable. The system in 
Table~\ref{tab:SPR_Lab} was modeled using these methods, and the dynamics of many 
ligand-receptor species with differing association and dissociation rates were simulated. 
The sensogram metrics defined in Table \ref{tab:metric} were employed to relate the 
mean-field approximation of the system to parameters easily extracted from the simulation 
data. 

The predictions of the sensogram metric were close to the actual simulation values for 
$Da<0.1$, but after that point the association rate begins to get large enough that 
diffusive transport begins to dominate the time scale on which ligands interact with 
receptors, and the probability of ligand rebinding events becomes very high. By ignoring 
these two temporal correlations, the mean-field predictions begin to drastically differ 
from the simulation parameters, and within a factor ten increase in the association rate, 
the error between the mean-field predictions and the simulation parameters increased by a 
factor of one hundred. Thus, these simulations show that a mean-field analysis of surface 
plasmon resonance is only valid for small values of $Da<0.1$, due to the importance of the 
diffusive and ligand-rebinding temporal correlations. Further work could be done on looking 
at the effects of the ligand-rebinding correlations on different receptor topologies. In 
biological systems, such as cells, receptors are not evenly distributed like those on the 
bottom of the SPR flow cell, but appear in clusters on the cell surface. This clustering 
could increase the likelihood that a ligand rebinding event occurs, allowing ligands to 
remain on the cell surface longer than would strictly be predicted from their binding rates,
c.f. Ref.~\cite{Tauber2}. This would further distance the dynamics of these biological 
systems from mean-field predictions.  

\ack
We gladly acknowledge helpful discussions with Michel Pleimling. 

\appendix 

\section{Reaction-diffusion-advection PDE}
This appendix is added to present a model of the SPR system described by 
Table~\ref{tab:SPR_Lab}, and to show that this can be reduced to a system of three
dimensionless parameters $Da$, $D_{D}$, $K$, and a time scale $\tau$.

The simplification of the advection-diffusion PDE follows from a derivation performed by 
Ref.~\cite{RT-Model}. We start with the PDE for ligand concentration in a flow cell with a 
receptor surface on the $y=0$ plane,
\begin{equation}
\label{eq:lig_flow_cell}
C_{t}=D(C_{xx}+C_{yy}) - (\frac{6v}{L_{y}^{2}})y(L_{y}-y)C_{x} \, ,
\end{equation}
where subscripts on {\it C} denote differentiation with respect to the subscript. 
Eq.~(\ref{eq:lig_flow_cell}) can be recast in terms of the scaled variables 
$\hat{x}=x/L_{x}$, $\hat{y}=y/L_{y}$, $\hat{z}=z/L_{z}$ and $\hat{t}=6vt/L_{x}$, 
\begin{equation}
\label{eq:dimless_lig_flow_cell}
C_{\hat{t}}=Pe^{-1}(\varepsilon^{2} C_{\hat{x}\hat{x}}+C_{\hat{y}\hat{y}})-
\hat{y}(1-\hat{y})C_{\hat{x}} \, ,
\end{equation}
where $\varepsilon=L_{y}/L_{x}$ is a dimensionless parameter, and $Pe$ denotes the P
ecl\'{e}t number
\begin{equation}
\label{ep:peclet_number}
Pe=\frac{6vL_{y}^{2}}{DL_{x}} \, ,
\end{equation}
which represents the ratio of the advective transport rate to the diffusive transport rate. 
The surface density of bound receptors ($R(\hat{x},\hat{t})$) evolves according to the 
reaction rate equation
\begin{equation}
\label{eq:receptor_rate}
R_{\hat{t}}(\hat{x},\hat{t})=k_{+}C(\hat{x},0,\hat{t})(R_{0}-R)-k_{-}R \, ,
\end{equation}
and the boundary condition for the receptor surface is given by 
\begin{equation}
\label{eq:concentration_boundary}
C_{\hat{y}}(\hat{x},0,\hat{t})=\frac{Pe}{L_{y}}R_{\hat{t}}(\hat{x},\hat{t}) \, .
\end{equation}
SPR systems typically have a Pecl\'{e}t number on the order of $100$. 

Now we can show that for systems with large Pecl\'{e}t numbers, close to the receptor 
surface (\ref{eq:dimless_lig_flow_cell}) simplifies and $Pe$ becomes irrelevant. First we 
redefine the $\hat{y}$ and $\hat{t}$ variables to a more useful form:
\begin{equation}
\label{eq:y_t_redefine}
\eta=Pe^{\alpha}\hat{y} \, , \quad \quad \quad \tau=Pe^{\beta}\hat{t} \, ,
\end{equation}
where $\alpha$ and $\beta$ are quantities that will be determined later. Using these 
substitutions, eq.~(\ref{eq:dimless_lig_flow_cell}) becomes
\begin{eqnarray}
\label{eq:redefined_lig_flow_cell}
C_{\tau}=&Pe^{-(\alpha+\beta)}(\varepsilon^{2}C_{\hat{x}\hat{x}}+Pe^{2\alpha}C_{\eta\eta})& 
\nonumber \\
&-(Pe^{-(\alpha+\beta)}\eta+Pe^{-(2\alpha+\beta)}\eta^{2})C_{\hat{x}} \, .&
\end{eqnarray}
If we require the P\'{e}clet coefficients on $C_{\eta\eta}$ and $\eta C_{\hat{x}}$ to be 
unity, the exponents $\alpha$ and $\beta$ must be $\alpha=1/3$ and $\beta=-1/3$. 
Eq.~(\ref{eq:redefined_lig_flow_cell}) then reduces to
\begin{equation}
\label{eq:exp_lig_flow}
C_{\tau}=Pe^{-2/3}\varepsilon^{2}C_{\hat{x}\hat{x}}+C_{\eta\eta} 
- (\eta-Pe^{-1/3}\eta^2)C_{\hat{x}}\, .
\end{equation} 
Because $\eta$ is a rescaling of $\hat{y}$, the only part of eq.~(\ref{eq:exp_lig_flow}) 
that determines the binding dynamics is the region where $\eta \rightarrow 0$. In this 
limit (\ref{eq:exp_lig_flow}) simplifies to
\begin{equation}
\label{eq:limit_exp_lig_flow}
C_{\tau}=Pe^{-2/3}\varepsilon^{2}C_{\hat{x}\hat{x}}+C_{\eta\eta} - \eta C_{\hat{x}} \, .
\end{equation}
Then, in the regime where $Pe^{-2/3}\varepsilon^{2}$ is small, the ligand concentration is 
governed by the reduced equation
\begin{equation}
\label{eq:limit2_exp_lig_flow}
C_{\tau}=C_{\eta\eta} - \eta C_{\hat{x}} \, .
\end{equation}
Finally, the ligand and receptor concentrations can be rendered dimensionless by the 
transformation
\begin{eqnarray}
&c(\hat{x},\eta,\tau)&=C(\hat{x}, \eta, \tau)/C_0 \, , \nonumber \\
&r(\hat{x},\eta,\tau)&=R(\hat{x}, \eta, \tau)/R_0 \, .
\end{eqnarray}
Under this transformation, the boundary conditions on the receptor surface given by 
eqs.~(\ref{eq:concentration_boundary}) and (\ref{eq:receptor_rate}) become
\begin{eqnarray}
& c_{\eta}(\hat{x},0,\tau)=& D_{D}^{-1}r_{\tau}(\hat{x},\tau) \, , \nonumber \\
& r_{\tau}(\hat{x},\tau)=& DaD_{D}\{c(\hat{x},0,\tau)(1-r)-Kr\} \, ,
\end{eqnarray}
where $Da$, $D_{D}$, $K$, and $\tau$ are defined in eqs.~(\ref{eq:tau})--(\ref{eq:K}).

\section{Scaling Method for Simulation Parameters}

Taking the laboratory parameters from Table~\ref{tab:SPR_Lab} and converting them into 
simulation parameters as listed in Table~\ref{tab:Model_Params} yields values too large to 
simulate in a reasonable amount of time. Therefore, it is necessary to find a method of 
scaling that can shrink this dynamical system down to an equivalent simulation cell. 

There are four parameters that characterize the system \cite{RT-Model}. These are derived 
in Appendix~A, and are summarized below. These are $\tau$, the time scale 
of the diffusive reactive system:
\begin{equation}
\label{eq:tau}
\tau = \Big(\frac{6v}{L_{x}}\Big)^{2/3}\Big(\frac{D}{L_{y}^2}\Big)^{1/3}t \, . 
\end{equation}
The Damk{\"o}hler number $Da$ is the ratio of the rate of ligand binding action at the 
receptor surface to the rate of transport to that surface:
\begin{equation}
\label{eq:Da}
Da = k_{+}R_{0}\Big(\frac{L_{x}L_{y}}{6vD^{2}}\Big)^{1/3}.
\end{equation}
$D_{D}$ is the ratio at which ligands diffuse across the vertical axis of the lattice, 
to the rate of transport to the receptors:
\begin{equation}
\label{eq:Dd}
D_{D}=\frac{C_{0}}{R_{0}}\Big(\frac{L_{x}L_{y}D}{6v}\Big)^{1/3}.
\end{equation}
Finally, $K$ represents the equilibrium dissociation constant for the reaction, 
normalized by the ligand concentration:
\begin{equation}
\label{eq:K}
K=\frac{k_{-}}{C_{0}k_{+}} \, .    
\end{equation}
\begin{figure}[h]
\centering
\includegraphics[width=.7\textwidth]{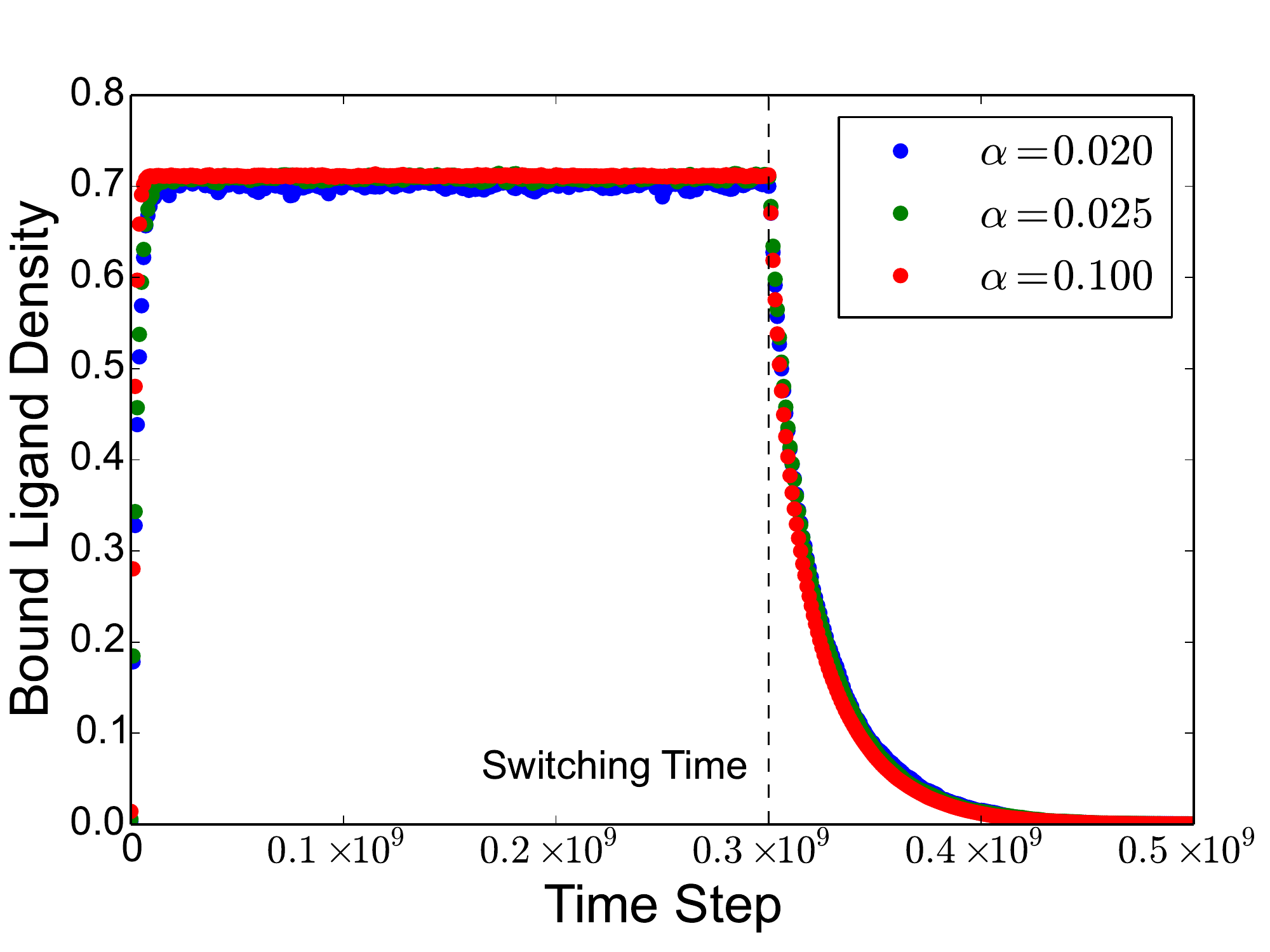}
\caption{An example of scaling for various values of $\alpha$. The density of bound ligands 
is plotted against the unscaled Monte Carlo time step for three realizations of the system 
described in Table~\ref{tab:Model_Params} with association rate $k_{+}=10^{6}M^{-1}s^{-1}$ 
and dissociation rate $k_{-}=10^{-2}s^{-1}$. The simulations were performed using three 
different values of the scaling parameter $\alpha$. Note that changing the values of 
$\alpha$ by a factor of $5$ implies a rescaling of the system length in the $x$ direction
and of the overall time scales by a factor of $25$. The results of these simulations were 
unscaled by multiplying by the reciprocal of the scaling factors when needed, and plotted 
versus the unscaled time steps. The unscaled concentration of ligands is 100nM for each 
simulation. This concentration is held constant for the duration of the association phase, 
which lasts until the $0.3\times 10^{9}$ time step. At this point, marked by the (black) 
dashed line and labeled as the `switching time', the concentration of incoming ligands is 
set to zero, to initiate the dissociation phase. The three data sets are represented by the 
(blue, red, and green) dots, and as expected, each of the three sets of data coincide.
\label{fig:scale}}
\end{figure}
Any method of scaling that preserves the dynamics of the system must keep these values 
unchanged. We may hence scale each of the physical parameters in these four values by a 
scale parameter $\alpha$ specified such that the values $Da$, $D_{D},$ and $K$ remain fixed: 
\begin{eqnarray*}
& L_{x} \rightarrow \alpha^{\gamma_{x}}L_{x} \, ,& \quad \quad  k_{+} \rightarrow 
  \alpha^{\gamma_{+}}k_{+} \, , \quad \quad v \rightarrow \alpha^{\gamma_{v}}v \, ,\\
& L_{y} \rightarrow \alpha^{\gamma_{y}}L_{y} \, ,& \quad \quad  k_{-} \rightarrow 
  \alpha^{\gamma_{-}}k_{-} \, , \quad \quad D \rightarrow \alpha^{\gamma_{D}}D \, ,\\
& C \rightarrow \alpha^{\gamma_{C}}C \, ,& \quad \quad  R \rightarrow 
  \alpha^{\gamma_{R}}R \, , \quad \quad \quad t \rightarrow \alpha^{\gamma_{t}}t \, ,
\end{eqnarray*}
where the constant $\alpha$ is a positive real number. We choose the exponents such that
\begin{eqnarray}
& 0 =& \gamma_{t} + \frac{1}{3}(\gamma_{D}+2\gamma_{v}-2\gamma_{x} - 2\gamma_{y}) \, , \nonumber \\
& 0 =& \gamma_{+}+\gamma_{R} + \frac{1}{3}(\gamma_{x}+\gamma_{y}-\gamma_{v}-2\gamma_{D}) \, , \nonumber\\
& 0 =&  \gamma_{C} -\gamma_{R} + \frac{1}{3}(\gamma_{x} +\gamma_{y} + \gamma_{D} - \gamma_{v})
\, , \nonumber\\
& 0 =& \gamma_{-} - \gamma_{C} - \gamma_{+} \, .
\end{eqnarray}
The above requirements ensure that none of the four parameters are affected by this scaling. 
At this point any exponents that satisfy the above requirements can be chosen. For 
simplicity's sake, the exponents of $v$, $D$, and $R$ were chosen to be zero. $\gamma_{y}$ 
and $\gamma_{x}$ were chosen to be $1$ and $2$ respectively. This yields the following 
definitions
\begin{eqnarray}
& \gamma_{x} = 2 \, ,& \quad \quad \quad \gamma_{t} = 2 \, , \quad \quad \quad \quad \gamma_{v} = 0 \, , \nonumber \\
& \gamma_{y} = 1 \, ,& \quad \quad \quad \gamma_{+} = -1 \, , \quad \quad \quad  \gamma_{D} = 0 \, , \nonumber \\
& \gamma_{C} = -1 \, ,& \quad \quad \quad \gamma_{-} = -2 \, ,\quad \quad \quad   \gamma_{R} = 0 \, .
\end{eqnarray}
Fig.~\ref{fig:scale} shows the results of simulations of the system described in 
Table~\ref{tab:Model_Params} with association rate $k_{+}=10^{6}M^{-1}s^{-1}$ and 
dissociation rate $k_{-}=10^{-2}s^{-1}$ scaled with various scaling constants $\alpha$. 
The range of values of $\alpha$ shown here is actually representative of a whole order of 
magnitude of values after $\alpha$ has been raised to the appropriate exponents.  
Note that the coincidence of the differently scaled simulation results confirms the 
assertion that the results of scaled simulations of the system described in 
Table~\ref{tab:Model_Params} will accurately represent the dynamics of the unscaled system.

\section{Algorithm for Monte Carlo Simulation}
A summary of the algorithm used for the Monte Carlo simulation is as follows.
\renewcommand{\labelenumi}{\arabic{enumi})}
\renewcommand{\labelenumii}{\alph{enumii})}
\renewcommand{\labelenumiii}{\roman{enumiii})}
\begin{enumerate}
\item Select a random ligand and generate a random number $r$ uniformly distributed between zero and one.
\item If the ligand is not bound to a receptor:
  \begin{enumerate}
  \item If $r < p_{0}$ the ligand remains at the same location.
  \item If instead $r <p_{0} + p_{x}^{+}$ the ligand is stepped in the positive $x$ direction.
    \begin{enumerate}
      \item If the ligand encounters the end of the SPR cell ($x = \widetilde{L_{x}}$), remove the ligand.
      \item If the simulation is in the association phase, introduce a new ligand at the $x=0$ plane to maintain ligand concentration.
    \end{enumerate}
  \item If instead $r <p_{0} + p_{x}^{+} + p_{x}^{-}$ the ligand is stepped in the negative $x$ direction.
    \begin{enumerate}
      \item If the ligand encounters the beginning of the SPR cell ($x = 0$),  do not move the ligand.
    \end{enumerate}
  \item If instead $r <p_{0} + p_{x}^{+} + p_{x}^{-} + p_{y}^{+}$, step the ligand in the positive $y$ direction. Otherwise if $r <p_{0} + p_{x}^{+} + p_{x}^{-} + p_{y}^{+} + p_{y}^{-}$, step the ligand in the negative $y$ direction.
    \begin{enumerate}
    \item If the ligand encounters either the top or bottom planes of the SPR cell (i.e. $y=0$ or $y=\widetilde{L_{y}}$), reflect the ligand back one lattice spacing into the lattice to ensure reflective boundary conditions.
    \end{enumerate}
  \item If instead $r <p_{0} + p_{x}^{+} + p_{x}^{-} + p_{y}^{+} + p_{z}^{+}$, step the ligand in the positive $z$ direction. Otherwise if $r <p_{0} + p_{x}^{+} + p_{x}^{-} + p_{y}^{+} + p_{y}^{-} +p_{z}^{+}+p_{z}^{-}$, step the ligand in the negative $z$ direction.
    \begin{enumerate}
    \item If the ligand moves past either of the $z$ axis boundaries of the SPR cell (i.e. $z=0$ or $z=\widetilde{L_{z}}$), place the ligand on the opposite boundary to create periodic boundary conditions.  
    \end{enumerate}
  \item After the ligand is stepped, if it is one lattice site above an empty receptor, generate a random number $q$ evenly distributed between zero and one.
    \begin{enumerate}
    \item If $q < \widetilde{k_{+}}$, bind ligand and receptor, and set ligand position to receptor position.
    \end{enumerate}
  \end{enumerate}
\item If the ligand is bound to a receptor, check if $r<\widetilde{k_{-}}$. If it is, unbind the ligand.
\item Repeat the above process $n$ times every time step, where $n=\widetilde{C_{0}}\cdot(\widetilde{L_{x}}\cdot \widetilde{L_{y}} \cdot \widetilde{L_{z}})$ is the number of ligands in the SPR cell.
\item Count the number of bound ligand receptor pairs and divide by the number of ligands in the volume of the SPR cell bounded on the bottom by the receptor surface during the association phase to retrieve the bound ligand density. Record this every time step.
\item After a steady-state concentration of bound ligand-receptor pairs has been reached, change from the association stage to the dissociation stage.
\end{enumerate}


\section*{References}

\begin{thebibliography}{10}
\bibitem{Biochem-Book}
Nelson D and Cox M 2004 {\it Lehninger Principles of Biochemistry, Fourth Edition} (New York: W.H. Freeman and Company)

\bibitem{Voet}
Voet D and Voet J 2011 {\it Biochemistry, Fourth Edition} (Hoboken, New Jersey: John Wiley \& Sons, Inc.)

\bibitem{SPR-Book}
de Mol N (ed.) and Fischer M (ed.) 2010 {\it Surface Plasmon Resonance} (Berlin: Springer-Verlag) 

\bibitem{Phizicky}
Phizicky E M and Fields S 1995 Protein-protein interactions: methods for detection and analysis. {\it Microbiological Reviews} {\bf 59}

\bibitem{Rich-1}
Rich R and Myszka D 2006 Survey of the year 2005 commercial optical biosensor literature {\it J. Mol. Recognit.} {\bf 19}.
\bibitem{Rich-2}
Rich R and Myszka D 2007 Survey of the year 2006 commercial optical biosensor literature {\it J. Mol. Recognit.} {\bf 20}
\bibitem{Rich-3}
Rich R and Myszka D 2008 Survey of the year 2007 commercial optical biosensor literature {\it J. Mol. Recognit.} {\bf 21}

\bibitem{Tauber}
Gopalakrishnan M, Forsten-Williams K, Cassino T, Padro L, Ryan T and T\"{a}uber U C 2005 Ligand rebinding: self-consistent mean-field theory and numerical simulations applied to surface plasmon resonance studies. {\it Eur Biophys J.} {\bf 34}

\bibitem{RT-Model}
Edwards D 1999 Estimating rate constants in a convection-diffusion system with a boundary reaction {\it IMA Journal of Applied Mathematics} {\bf 63}

\bibitem{My1}
Myszka D G, Morton T A, Doyle M L and Chaiken I M 1997 {\it Biophys. Chem.} {\bf 64}
\bibitem{My2}
Myszka D G, He X, Dembo M, Morton T A and Goldstein B 1998 Extending the range of rate constants available from BIACORE: interpreting mass transport-influenced binding data {\it Biophys. J.} {\bf 75}

\bibitem{PDE-Sim}
Hu G, Gao Y and Li D 2007 Modeling micropatterned antigen–antibody binding kinetics in a microfluidic chip {\it Biosensors and Bioelectronics} {\bf 22}

\bibitem{Monte_Carlo}
Schnoerr D, Sanguinetti G and Grima R 2016 Approximation and inference methods for stochastic biochemical kinetics - a tutorial review {\it e-print} 
{\tt arXiv:1608.06582}

\bibitem{Tauber2}
Gopalakrishnan M, Forsten-Williams K, Nugent M A and T\"{a}uber U C 2005 Effects of Receptor Clustering on Ligand Dissociation Kinetics: Theory and Simulations {\it Biophysical Journal} {\bf 89}

\bibitem{Rate-Equation}
Motulsky H and Mahan L 2014 The Kinetics of Competitive Radioligand Binding Predicted by the Law of Mass Action {\it Molecular Pharmacology} {\bf 86}

\bibitem{Schasfoort}
Schasfoort R (ed.) and Tudos A (ed.) 2008 {\it Handbook of Surface Plasmon Resonance} (Cambridge: The Royal Society of Chemistry)

\bibitem{SPR_Procedure}
Zeng S, Yu X, Law W, Zhang Y, Hu R, Dinh X, Ho H and Yong 2013 Size dependence of Au NP-enhanced surface plasmon resonance based on differential phase measurement. {\it Sensors and Actuators B: Chemical.} {\bf 176}

\bibitem{RU-scaling}
Davis T and Wilson W 2000 Determination of the refractive index increments of small molecules for correction of surface plasmon resonance data {\it Analytical Biochemistry} {\bf 284}

\bibitem{reynolds_number}
Zourob M (ed.), Elwary S (ed.) and Turner A (ed.) 2008 {\it Principles of Bacterial Detection: Biosensors, Recognition Receptors and Microsystems} (New York: Springer). 

\bibitem{LL-Fluid}
Landau, L D and Lifshitz E M 1998 {\it Fluid Mechanics}
(Oxford: Butterworth-Heinemann), second edition

\bibitem{Papalia-Rates}
Papalia G, Leavitt S,  Bynum M, Katsamba P, Wilton R,
 Qiu H,  Steukers M, Wang S,  Bindu L,  Phogat S, Giannetti A, Ryan T, et al. 2006 Comparative analysis of 10 small molecules binding to carbonic anhydrase II by different investigators using Biacore technology {\it Analytical Biochemistry} {\bf 359}

\bibitem{LL-Rates}
Lauffenburger D and Linderman J 1993 {\it Receptors. Models
for Binding, Trafficking, and Signaling.} (New York: Oxford University
Press)


\bibitem{Glaser}
Glaser R W 1993 Antigen-Antibody Binding and Mass Transport by Convection and Diffusion to a Surface: A Two-Dimensional Computer Model of Binding and Dissociation Kinetics {\it Analytical Biochemistry} {\bf 213}

\bibitem{Schuck-Minton}
Schuck P and Minton A 1996 Analysis of Mass Transport-Limited Binding Kinetics in Evanescent Wave Biosensors {\it Analytical Biochemistry} {\bf 240}

\bibitem{receptors}
Oliver J M and Berlin R 1982 Distribution of receptors and functions on cell surfaces: Quantitation of ligand-receptor mobility and a new model for the control of plasma membrane topography {\it Philosophical Transactions of the Royal Society of London. B, Biological Sciences} {\bf 299}

\end{thebibliography}
\end{document}